\documentclass[prb,preprint,superscriptaddress,showpacs]{revtex4-1}
\usepackage{amsmath}
\usepackage{amssymb}
\usepackage{graphicx}
\usepackage[utf8]{inputenc} 
\usepackage[T1]{fontenc}      
\usepackage{textcomp}            

\usepackage[abs]{overpic}

\usepackage[hypertex]{hyperref} 

\def\atanh{{\rm atanh}}
\def\st{{\vec \sigma}}
\begin{document}

\title[The cavity method for quantum disordered systems]{The cavity method for quantum disordered systems: from transverse
  random field ferromagnets to directed polymers in random media}
\author{ O. Dimitrova and M. M\'{e}zard}
\address{
Laboratoire de Physique Th\'eorique et Mod\`eles Statistiques,
CNRS and Universit\'e Paris-Sud, B\^{a}t 100, 91405 Orsay Cedex, France}

\begin{abstract}
After reviewing the basics of the cavity method in classical systems,
we show how its quantum version, with some
appropriate approximation scheme, can be used to study a system of
spins with random ferromagnetic interactions and a random transverse
field. The quantum cavity equations describing the
ferromagnetic-paramagnetic phase transition can be transformed into
the well-known 
problem of a classical directed polymer in a random medium. The glass
transition
of this polymer problem translates into the existence of a
`Griffiths phase' close to the quantum phase transition of the quantum
spin problem, where the physics is dominated by rare events. The
physical behaviour of random transverse field ferromagnets on the
Bethe lattice is found to be very similar to the one found in finite
dimensional systems, and the quantum cavity method gets back the known
exact results of the one-dimensional problem.
\end{abstract}

\maketitle
\section{Introduction}
The cavity method has been developed to study classical frustrated spin
systems: spin glasses. In recent years its range of application has
broadened a lot as it was applied with some success to hard computer
science problems and also to some quantum problems. This paper
aims at giving some background on the cavity method, both classical
and quantum, and explaining how it can be used
to study a quantum problem, a  ferromagnetic spin
system in a random transverse field, following the recent work of~[\cite{IoffeMezard2010,FeiIofMez2010}]. This is a problem which has been
studied a lot in one dimension; we analyze  it here using the cavity
method which is a mean-field type method better suited for large
dimensional systems. 
However we find that the physics is
rather similar to the one found in the one-dimensional
case, in the sense that rare events (rare sites with anomalously small
transverse fields) play a major role in
the neighborhood of the quantum critical point. This leads to very
large spatial fluctuations of the spontaneous magnetization in the
ferromagnetic phase, and a broad distribution of the local magnetic
susceptibilities in the paramagnetic phase. These effects, which are
typical of the `Griffiths phase' described in low dimensions, are found
here through the use of an auxiliary problem of directed polymers
in random media: the rare-event dominated regime of the
ferromagnetic-paramagnetic phase transition is related to the glass
transition
of the polymer problem.

This paper is organized as follows.
Sec.\ref{sec:ccav} gives an introduction to the cavity method in
classical spin systems. Sec.\ref{rtff_def} defines the random
transverse field ferromagnet and gives some background on this
problem.
The naive mean field approach to this problem is explained in  Sec.\ref{se:MF}.
Sec. \ref{sec:qcav} introduces the quantum cavity method and
discusses some systematic approximation schemes which can be used in
this context. These approximation schemes are tested in
Sec.\ref{se:ferro} by applying them to a very simple, and very well
understood, problem: the pure ferromagnet in a transverse field.
Sec.\ref{RTFF} describes the use of the quantum cavity method, with a
`cavity-mean-field' approximation, to study the phase diagram of a
ferromagnet in a random transverse field. It derives the phase transition line in
the plane disorder-temperature, and describes the main properties of
the low-temperature ferromagnetic and paramagnetic phases, in
particular the large fluctuations of the order parameter and of the
susceptibility. This section ends by a study of the one-dimensional
case, where the cavity method gets the known exact
results, and a discussion of the higher-dimensional cases, for which
the Bethe lattice analysis provides a useful mean-field approximation. 
A short summary  is given in Sec.\ref{se:conc}.

\section{The classical cavity method}
\label{sec:ccav}
\subsection{The Sherrington-Kirkpatrick model}
The cavity method has been introduced twenty five years ago in the context of spin glass mean field theory~\cite{MPV_cav}.
The solution of the fully connected Sherrington-Kirkpatrick (SK)~\cite{SK} model had been found a few years before by G. Parisi~\cite{GP_RSB}, who had proposed,
within the replica approach, 
an inspired Ansatz of replica symmetry breaking (RSB). The cavity method gave an alternative solution, with identical physical content and 
results to the replica approach, but which was much more transparent. Instead of the mysterious replicas, it used a purely probabilistic approach,
 based on three assumptions which had been found to be hidden in the form of Parisi's RSB Ansatz: the existence of many pure states~\cite{Parisi83},
 ultrametricity~\cite{MPSTV85}, and the exponential distribution of free energies of the pure states~\cite{MPV85}. This approach has turned out to
 be very powerful and has served as a starting point of recent works by Talagrand~\cite{Talag} and Guerra~\cite{Guerra} which established rigorously 
the exactness of the free energy obtained by the RSB/cavity method (although the ultrametric property has still not be proven).

In a nutshell, the cavity method as it was applied to the SK model consists in assuming some structure for a $N$-spin system $s_1,\dots,s_N$, 
and checking self consistently that this structure is reproduced when one adds a new spin $s_0$ and goes to a $N+1$ spin-system. Within the
 simplest replica symmetric (RS) approximation, the main assumption is that the correlation functions are small. The local field on $s_0$ then has
 a Gaussian distribution, with a width which must be reproduced
 self-consistently and gives the Edwards-Anderson order parameter~\cite{EA}.
 This approach gives the RS solution, which is correct at temperatures
 above the de-Almeida Thouless (AT) line~\cite{AT}. It breaks down in
 the
 spin glass phase,  when the proliferation of pure states induces a
 non-trivial correlation between distinct spins. In order to study the
 spin glass 
phase, one must explicitly assume the existence of many pure states~\cite{TAP,BrayMoore} with free energies distributed 
as a Poisson process with exponential distribution~\cite{MPV_cav}. Within each state,
the correlations are small and the cavity method 
can be applied. But the addition of the new spin $s_0$ creates a free
energy shift which depends on the state:
 the free energies are reshuffled, and the selection of low free
 energy states (by the Boltzmann weight) biases
 the local field distribution. The whole cavity RSB approach relies on
 controlling these crossings of the free
 energies of the states, and checking the self consistency of the
 basic hypotheses. The results are exactly
 identical to those of the replica approach.

\subsection{Finite connectivity spin glasses}
\label{subsec:diluted}
\begin{figure}
 \centering
 \includegraphics[scale=0.2]{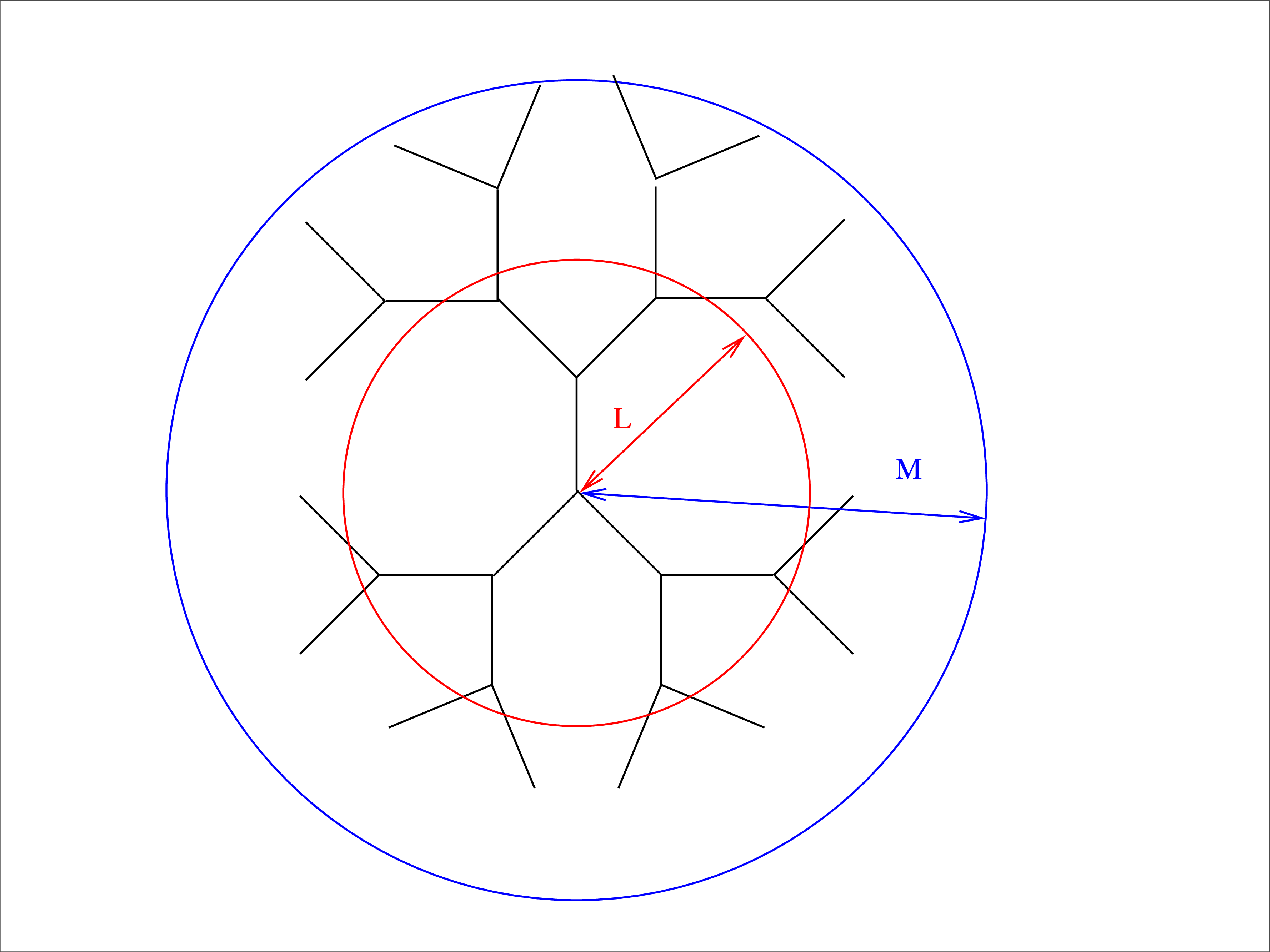}
 \includegraphics[scale=0.4]{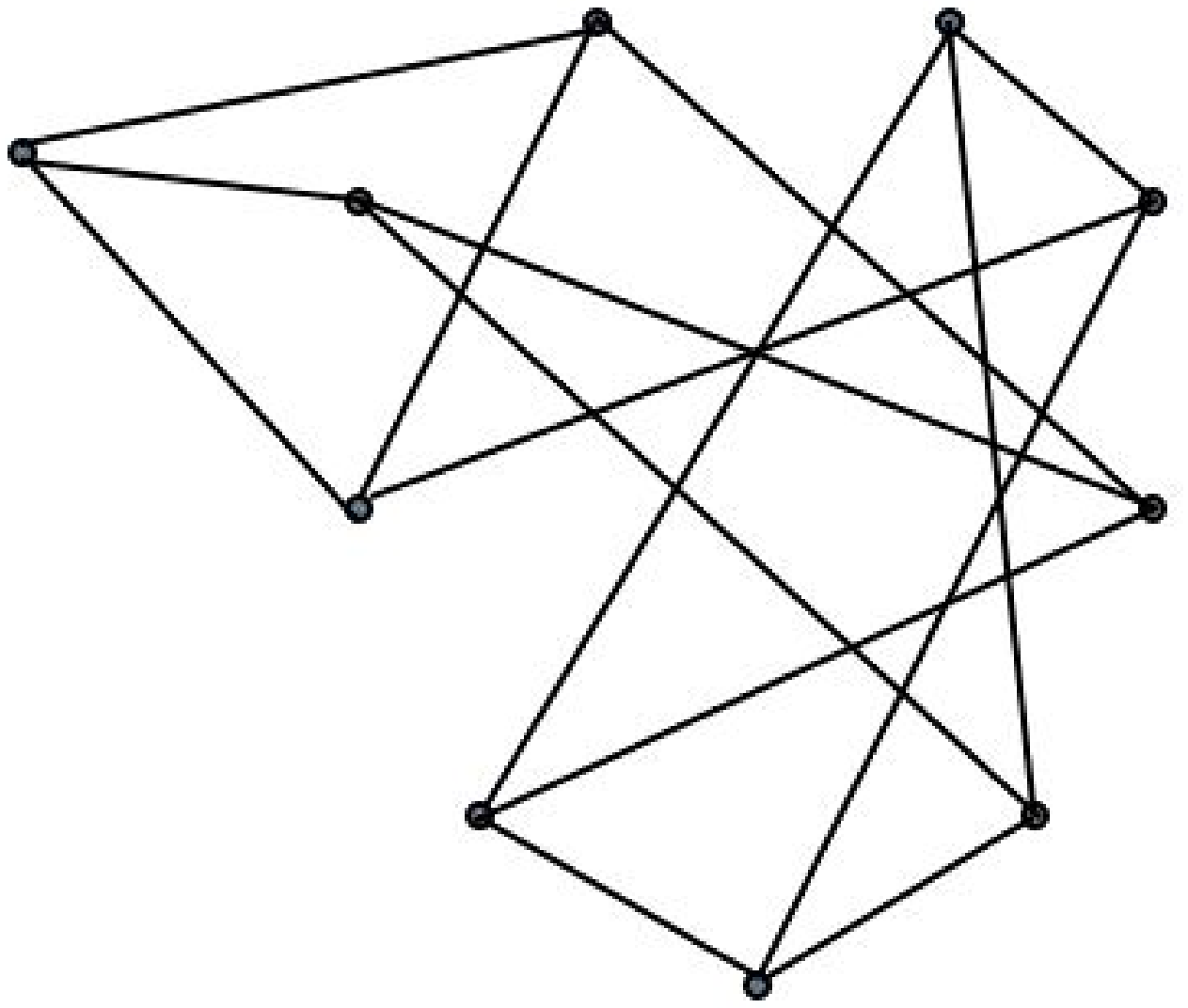}
  \caption{Left: the usual construction of the Bethe lattice. Right: a (small) random regular graph. The local structure in large random regular graph is tree-like, as in the center of a Bethe lattice.}
\label{fig:Bethe}
\end{figure}

It was soon realized that it would be very instructive to go beyond the SK model by studying spin glass models in which each spin has only a finite number of neighbours. In order to introduce mean field models with this property, the usual approach is to study the Bethe lattice, which is usually defined as the interior of a Cayley tree: taking a large Cayley tree of depth $M$, one studies the interior part up to depth $L$. The Bethe lattice is then defined by the double limit $\lim_{L\to\infty}\lim_{M\to\infty} $(see Fig. \ref{fig:Bethe}).
This procedure is fine for ferromagnetic interactions, but it raises a
problem when one studies spin glasses where a crucial ingredient is
the frustration due to loops in which the product of exchange couplings is negative.
A simple procedure is to induce frustration by fixing randomly (or
imposing some type of local fields on) the spins at the boundary of
the tree. However most of the physics is then put by hand through this
boundary condition, which involves a finite fraction of the total
number of spins. For instance fixing the boundary spins to $\pm 1$
independently with probability 1/2~\cite{thouless,chayes} gives a
system which does not have a genuine spin glass phase (technically it
is  is always replica symmetric). It has been found recently how the boundary conditions should be fixed in order to get RSB: the correct procedure involves the process of broadcast and reconstruction~\cite{MezMon_broadcast} and creates  subtly correlated boundary fields.

An alternative construction is to define the Bethe lattice spin glass through a random regular graph, taken uniformly from all the graphs where each
 spin has $K+1$ neighbours (see Fig. \ref{fig:Bethe}). Large random regular graphs with $K\gg 1$ vertices have a locally tree-like structure like 
the interior of a Cayley tree, as the typical size of a loop is of order $\log N/\log K$. These large loops create the frustration without having to 
play with boundary conditions. The study of spin glasses on random
graph with a finite connectivity started with Viana and Bray~\cite{VianaBray}. The  solution of spin glasses on random regular graphs
within the RS approximation was found relatively early~\cite{VianaBray,MezPar87,KanterSompo87}, but
it took fifteen more years to understand how to handle RSB effects~\cite{MPBethe},
 and this has been done so far only at the level of one, or at most
 two, steps of RSB. Here we shall briefly review how the cavity method
 can be used to study this problem, mostly at the RS level, as this gives the building blocks of the quantum cavity method which we use below.

\begin{figure}
 \centering
 \includegraphics[scale=0.4]{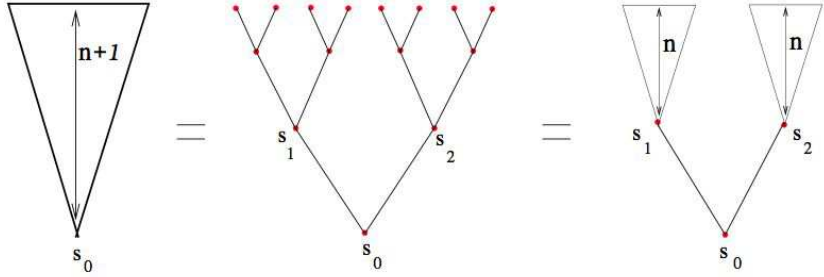}
  \caption{The recursion relation for the local magnetic field on the root of a rooted tree, here for the case $K=2$.}
\label{fig:BPrec}
\end{figure}

Let us study the equilibrium properties, at inverse temperature $\beta$, of a  general Ising spin glass problem with a random field on a Bethe lattice, defined by the Hamiltonian
\begin{equation}
H=-\sum_{<ij>} J_{ij} s_i s_j -\sum_i B_i s_i
\end{equation}
where $<ij>$ are the links of a random-regular graph of degree $K+1$. The idea is to use the local tree-like property for rooted trees. Let us pick up 
one spin, say $s_0$, and delete the edge between $0$ and one of its neighbours. Up to any finite distance, $s_0$ is now the root of a tree. Let us denote by $h_0$ the local 'cavity' 
magnetic field on the root $s_0$ due to this tree, so that $\langle s_0\rangle=\tanh(\beta h_0)$. Denote by  $s_1,\dots, s_K$ its neighbours on the tree, and by $h_1,\dots, h_k$ their cavity fields (so that for instance the magnetization of $s_1$ in the absence of $s_0$ is $\tanh(\beta h_1)$). The Boltzmann measure on $s_0$ is then
\begin{equation}
P(s_0)= \frac{1}{Z_0} \sum_{\{s_i\}}\exp\left(\beta\left[B_0 s_0+\sum_{i=I}^K (J_{0i} s_0 s_i+h_i s_i)\right]\right)
\end{equation}
By doing explicitely the sum over the spins $s_i$, one gets the recursion
\begin{equation}
h_0=B_0+\frac{1}{\beta}\sum_{i=1}^K \atanh(\tanh(\beta
J_{0i})\tanh(\beta h_i))
\label{hupdate}
\end{equation}
which is illustrated pictorially in Fig.\ref{fig:BPrec}. One way to describe it is to build the tree rooted in $s_0$ by considering the 
$K$ trees rooted on $s_1,\dots, s_K$ (in absence of $s_0$) and merging them.
These recursion `cavity' equations are always correct on a tree. On a random regular graph they are correct whenever the joint probability of the spins $s_1,\dots,s_K$
in the absence of $s_0$ factorizes into a product $\prod_i \exp\left(\beta h_i s_i\right)/(2\cosh(\beta h_i))$. Two conditions are needed for this
 absence of correlations of neighbouring spins. One is the local tree-like property which guarantees that any two neighbouring spins $s_i,s_j$,
 with $i,j\in\{1,\dots,K\}$, are far away on the graph obtained by eliminating the site $s_0$ and all edges connected to it. The second one is that
 the correlations decay at large distance, so that the correlations
 between two neighbours  $s_i$ and $s_j$ (which are typically at distance $\log N/\log K$ when
 $s_0$ is absent), vanish in the large $N$  limit. This is true when the system is in its paramagnetic phase. It is not correct in a ferromagnetic
 phase or in  a spin glass phase because long-range correlations develop. However, if one is able to restrict the measure to one single pure state (where by definition correlations decay at large distance), then the recursion relations are correct. 

These recursion relations can be used in two main ways:

1- On a given
sample, one can see these equations as a set of self-consistent
equations relating cavity fields: there are two such cavity fields for each
edge $<ij>$ of the graph, one is the field on site $i$ in the absence
of $j$ (case where $i$ has been chosen as root), the other one is the field on site $j$ in absence of $i$.  All
these fields are related by the cavity equations. In
the large $K$ limit these equations reduce
to the 'TAP' equations~\cite{TAP}.  These RS cavity equations are
also well known in computer science under the name of `belief
propagation' equations~\cite{MezMon09}.

2- One can also study the statistics
of cavity fields. Suppose that the local magnetic fields $B_i$ are
independent random variables drawn from a distribution $P_{ext}(B)$, and the exchange
couplings are independent random variables drawn from a distribution $\rho(J)$. At the RS level, one expects a single solution for
all cavity fields on a
given instance. Then one can define the distribution of cavity
fields $P(h)$  when one picks up an oriented edge randomly. This is
the natural order parameter for this problem. Its knowledge provides
the full solution of the problem when the RS hypothesis is correct.
Technically, finding $P(h)$ is usually done by a `population dynamics'
method first introduced in~[\cite{Abou}], and developed in the present
context by~[\cite{MPBethe}]. Its idea is the following.
The update equation (\ref{hupdate}) induces a self-consistency
equation for $P(h)$ which can be written as:
\begin{equation}
P(h)=\int dB  P_{ext}(B)\; \prod_{i=1}^K \int dh_i P(h_i) dJ_i \rho(J_i)
\delta\left(
h-B-\frac{1}{\beta}\sum_{i=1}^K \atanh(\tanh(\beta
J_{i})\tanh(\beta h_i))
\right)
\end{equation}
The population dynamics method represents $P(h)$ by a population of
$M\gg 1$ fields $h_1,\dots,h_M$. This population is updated by a
Monte-Carlo-type process in which, at each iteration, one does the
following operations:
\begin{itemize}
\item
Choose $K$ indices $r_1,\dots r_K\in\{1,\dots, M\}$ randomly
uniformly.
\item 
Generate $K$ independent couplings $J_1,\dots J_K$ from $\rho(J)$, and
a field $B$ from $P_{ext}(B)$.
\item 
Compute $h=B+\frac{1}{\beta}\sum_{i=1}^K \atanh(\tanh(\beta
J_{i})\tanh(\beta h_{r_i})$
\item 
Choose an index $j\in\{1,\dots, M\}$ randomly
uniformly, and replace in the population the value $h_j$ by the
new value $h$.
\end{itemize}
The convergence of this method when the number of iterations is large
enough can be checked by monitoring for
instance moments of $P(h)$ (the $r$-th moment is evaluated as $(1/M)
\sum_{i=1}^Mh_i^r$).  
If $M$ is large enough the population will give a good approximation
of $P(h)$.

Note that the RSB situation is much
more complicated: on a given instance there are many solutions to the
system of equations relating the cavity fields. On a given oriented
edge (say looking at the cavity field on $i$ in absence of $j$) the
cavity field can take many values depending on the solution one
takes. Doing the statistics over all solutions defines a distribution
of fields on each oriented edge, and the order parameter is the
distribution of these distributions when one picks an edge at
random. In the following quantum problems we shall use only the RS version.

\section{Random transverse-field ferromagnets}
\label{rtff_def}
The problem of ferromagnets in random transverse fields has received
a lot of attention in recent years. These systems provide relatively simple examples of
 disordered systems displaying a quantum phase transition~\cite{Sachdev2000}. They are described by
the Hamiltonian:
\begin{equation}
H_{F}=-\sum_{i}\xi_{i}\sigma_{i}^{z}-\sum_{(ij)}J_{ij}\sigma_{i}^{x}\sigma_{j}^{x}\ .\label{H_B}
\end{equation}
where  $\sigma^z,\sigma^x$ are Pauli matrices. The exchange couplings
are independent ferromagnetic interactions drawn from
a distribution $\rho(J)$ which has support on $J> 0$, and the random
transverse fields are independent variables drawn from a distribution
$\pi(\xi)$. The lattice structure is described by the set of pairs
$(ij)$ which appear in the above sum; we suppose that this lattice is
homogeneous, and each spin has exactly $z=K+1$ neighbours. For definiteness we shall use a model where the
distribution of couplings and fields are given by:
\begin{equation}
\rho(J)=\frac{K}{2} \theta(J)\theta(2/K-J)\ \ \ ; \ \ \
\pi(\xi)=\frac{1}{h}\theta(\xi)\theta(h-\xi)\ ,
\end{equation}
where $\theta$ is the Heaviside function.
The choice of the width of the $J$ distribution  fixes the energy scale such that the mean-field critical temperature in
absence of disorder has $T_c=1$.  The Hamiltonian $H_F$ thus depends on a single parameter $h$
which measures the degree of disorder.

The one dimensional case, which is intimately related to the
two-dimensional classical model of MacCoy and Wu~\cite{MacCoyWu}, has been solved in great detail by
Fisher~\cite{Fisher1992} using the strong disorder Ma-Dasgupta-Hu
decimation procedure~\cite{MaDasgupta}. This solution has emphasized the
importance of Griffiths singularities, which manifest themselves most
notably in the fact that the average susceptibility differs from the
typical one. The method itself has a broad range of applications~\cite{Monthus}. These results have been confirmed
numerically~\cite{RiegerYoung,IgloiRieger98}, and numerical simulations of the same
problem in dimension two, both
through the strong disorder decimation and through Monte-Carlo simulations, also point to the same kind of physics as in
one dimension~\cite{Pich,RiegerKawashima,Motrunich,saleur,KovacsIgloi_prb2010}. Recent
results also find the same behaviour in dimension three and higher~\cite{KovacsIgloi10}

 Using the method introduced in~[\cite{IoffeMezard2010,FeiIofMez2010}], we shall now study the Bethe
approximation
for the
random transverse-field ferromagnet described by (\ref{H_B}). We thus
assume that the spins are on the vertices of a random regular graph of
degree $z=K+1$ and we want to study the phase diagram of this problem 
as function of the temperature $T$ and the amount of disorder $h$.

\section{Mean field}
\label{se:MF}
\subsection{Naive mean field}

For large $z$ one may be tempted to use the mean-field approach, in which $H_F$ is replaced
by $H_{MF}=\sum_{i}(-\xi_{i}\sigma_{i}^{z}-B\sigma_{i}^{x})$ where
$B$ is determined self-consistently from the equation $B=\sum_{j}\langle\sigma_{j}^{x}\rangle$.
At temperature
$T=1/\beta$,  the self consistent equation for $B$ obtained at $z\to
\infty$ is:
\begin{equation}
B=z\int d\xi\,
\pi(\xi)\,\int dJ\rho(J) \frac{B}{\sqrt{\xi^{2}+B^{2}}}\tanh(\beta\sqrt{\xi^{2}+B^{2}})\
.
\label{eq:BCS_equation}
\end{equation}
At finite temperature there is a transition between a large disorder
paramagnetic phase where $B=0$ is the only solution of this equation, to
a ferromagnetic phase characterized by a non-zero field, $B>0$. The
critical temperature $T_c$ is given by:
\begin{equation}
\frac{1}{h}\int_{0}^{h}d\xi\frac{\tanh(\xi/T_{c})}{\xi}=1\ .\label{Tc_MF}\end{equation}
At zero temperature this naive mean field approach predicts that the system
is always in its ferromagnetic phase.

While this approach is correct when $z=\infty$, its conclusions are qualitatively
wrong for any finite $z$.

\subsection{Bethe-Peierls approximation}
The naive mean field prediction that the system is always ferromagnetic at zero $T$
is wrong; the reason is that the mean field approach does not take
into account properly the rare events. Rare sites $j$ where the energy
$|\xi_j|\ll 1$ can be easily polarized in the $x$ direction, and they
play a key role in the establishment of the ferromagnetic
order. This effect, which has been first discussed qualitatively in~[\cite{MaHalperinLee}], leads to a zero temperature quantum phase
transition
to a paramagnetic state at a critical value of disorder $h_c$.
In order to study this quantum transition in details, we shall use the quantum version of the
cavity method.

\section{The quantum cavity method and its projections }
\label{sec:qcav}
\subsection{The RS quantum cavity method}
Let us study the transverse-field Ising Hamiltonian
(\ref{H_B})
on a Bethe lattice defined as a random regular graph.
The partition function $Z=Tr\ e^{-\beta H_{I}}$can be expressed with
the Suzuki-Trotter representation. Using $M$ imaginary time steps,
we introduce at each time step a decomposition on the eigenvectors of
the operators $\sigma_j^x$.
This describes the system by  the classical time trajectory of each spin, $\sigma_{i}(t)\in{\pm
  1}$
, where $t=1,\dots,M$, and $\sigma_j(t)$ denotes the eigenvalue of
$\sigma_j^x$ at time $t$. Then 
\begin{equation}
Z=\lim_{M\to\infty}\ \sum_{\left\{ \sigma_{i}(t)\right\} }e^{-\beta H_{ST}}
\end{equation}
where:
\begin{equation}
H_{ST}=-\frac{1}{M}\sum_{t}\sum_{(ij)}J_{ij}\sigma_{i}(t)\sigma_{j}(t)-\sum_{t}\sum_{i}\Gamma_{i}\sigma_{i}(t)\sigma_{i}(t+1)
\end{equation}
and $\Gamma_{i}=\frac{1}{2\beta}\log\coth\left(\frac{\beta\xi_{i}}{M}\right)$

This Hamiltonian acts  on spin trajectories
$\st_j=\{\sigma_{i}(t),\ t=1,\dots,M\}$. Each trajectory 
can be seen as a variable $\st_j $ taking $2^M$ possible values, and $H_{ST}$
defines the interaction of these $N$ variables. The crucial point is
that this Hamiltonian involves interactions between pairs of
trajectories which are neighbours on the Bethe lattice. This is
a locally tree-like graph and therefore one can use the cavity method
to study it. 
The RS cavity method for this problem was introduced in~[\cite{Laumann}]. It is obtained from the classical cavity
method described in Sec.\ref{subsec:diluted} by using spin trajectories
instead of Ising spins.
Take a branch of the Bethe lattice rooted on spin $0$, and define
the probability distribution of the time-trajectory of this spin as
$\psi_{0}\left[\st_{0}\right]$. Define analogously,
for each of the $K$ spins $ i$ which are the first neighbors of
$ 0$ on the rooted tree, their time trajectory as $\psi_{i}\left[\st_{i}\right]$.
Then one can write the mapping that generates, from $\left\{
  \psi_{i}\left[\st_i\right]\right\} $,
the new $\psi_{0}\left[\st_0\right]$:
\begin{eqnarray}\nonumber
\psi_{0}\left[\st_0\right]&=&C\sum_{\st_1,\dots,\st_K}
\psi_{1}\left[\st_1\right]\dots \psi_{K}\left[\st_K\right]\\ &&
\exp\left(\frac{\beta}{M}\sum_{t}\sum_{(i)}J_{0i}\sigma_{0}(t)\sigma_{i}(t)+\beta
  \sum_{t}\Gamma_{0}\sigma_{0}(t)\sigma_{0}(t+1)\right)
\label{st_update}
\end{eqnarray}
The population dynamics method of Sect.\ref{subsec:diluted} can be
used to study this cavity recursion: one must represent each of the $ \psi_{i}$ by a sample
of spin trajectories. This has been done for some problems in~[\cite{Laumann,Krzakala}],
but this approach tends to be rather heavy numerically. In
particular, in a disordered system like the random transverse-field
ferromagnet, the natural order parameter is the distribution of the
functions $\psi_j$ when the site $j$ is drawn randomly. This is a
rather complicated object which must be represented as a population
of populations.

\subsection{The projected cavity mapping}
While the exact RS cavity mapping is in principle doable, the
difficulty in its numerical resolution may make it difficult to get
a clear understanding of the physical results. For this reason one may
want to develop approximate versions of it. One such approximation,
the projected cavity mapping
introduced in~[\cite{IoffeMezard2010}], consists in using a local distribution
$\psi_{i}\left[\st_{i}\right]$ which takes the special form,
parameterized by one single number $B_i$:
\begin{equation}
\psi_{i}\left[\sigma_{i}\left(t\right)\right]=C\exp\left(\frac{\beta
    B_{i}}{M}\sum_{t}\sigma_{i}(t)+\beta\Gamma_{i}\sum_{t}\sigma_{i}(t)\sigma_{i}(t+1)\right)
\ .
\label{psi_proj}
\end{equation}
Injecting this form into (\ref{st_update}) we obtain a $\psi_0$ which
does not have the form (\ref{psi_proj}). However one can project it
back on a $\psi_0$ in the right subspace described by (\ref{psi_proj})
as follows. From the obtained
$\psi_0$ one can compute $\langle \sigma^x_0\rangle$ and deduce from
it the effective parameter $B_0$ such that the distribution $C\exp\left(\frac{\beta
    B_{0}}{M}\sum_{t}\sigma_{0}(t)+\beta\Gamma_{0}\sum_{t}\sigma_{0}(t)\sigma_{0}(t+1)\right)$
gives this value of $\langle \sigma^x_0\rangle$.

This projected cavity mapping is actually more easily understood
directly in terms of quantum Hamiltonians, without going to the
Suzuki-Trotter
formalism.  One studies the properties
of a spin $0$ in the rooted  graph where one of its neighbors has
been deleted, assuming that the $K$ remaining neighbors are
uncorrelated.
Each of these neighbours, when it is at the root of the subtree
obtained  by deleting the link to spin $0$, is supposed to be
described by a Hamiltonian

\begin{equation}
H_{i}^{cav}=-\xi_{i}\sigma_{i}^{z}-B_{i}\sigma_{i}^{x}
\end{equation}
The system of spin $0$ and its $K$ neighbors is thus described by
the local Hamiltonian

\begin{equation}
H_{0}=-\xi_{0}\sigma_{0}^{z}-\sum_{i=1}^{K}\left(\xi_{i}\sigma_{i}^{z}+B_{i}\sigma_{i}^{x}+J_{0i}
  \sigma_{0}^{x}\sigma_{i}^{x} \right)\label{eq:H^cav}\end{equation}
 By diagonalizing this $2^{K+1}\times 2^{K+1}$ Hamiltonian,
one can compute the induced magnetization of spin $0$,
$m_0=\langle\sigma_{j}^{x}\rangle$.
In order for this approach to be self-consistent, this magnetization
should be equal to the one obtained from the Hamiltonian
$ -\xi_{0}\sigma_{i}^{z}-B_{0}\sigma_{i}^{x}$. This means that $B_0$
is obtained by solving the equation:
\begin{equation}
\frac{B_{0}}{\sqrt{\xi_{0}^{2}+B_{0}^{2}}}\tanh\beta\sqrt{\xi_{0}^{2}+B_{0}^{2}}=m_0\
\end{equation}
The projected cavity mapping thus uses a single number $B_i$ to
describe the trajectory distribution $\psi_i$. It leads to a mapping
which gives  the new cavity field $B_{0}$
in terms of the $K$ fields $B_{i}$ on the neighboring spins. This
mapping induces a self-consistent equation for the distribution of
the $B$ fields, which is the natural order parameter in this
context: the paramagnetic phase has $P(B)=\delta(B)$, and the
ferromagnetic phase is described by a non-trivial $P(B)$.
The population dynamics method can be used to find this distribution.

\subsection{The cavity-mean-field approximation}
The projected cavity mapping is numerically much simpler than the full RS cavity
method using spin trajectories. It also allows to address a broader
range of questions, like those related to the real time dynamics and
the spin relaxation, which are not easily accessible within the
Suzuki-Trotter formalism~\cite{IoffeMezard2010,FeiIofMez2010}. Still
its practical use is a bit slow numerically as one must diagonalize
the $K+1$ spin Hamiltonian (\ref{eq:H^cav}) each time one wants to
generate  a new
field $B_0$ in the population.

It has turned out useful~\cite{IoffeMezard2010} to use one more step of approximation in
order to obtain an explicit mapping, similar to the one found in the
classical problem (\ref{hupdate}). This can be done using a mean field
approximation in order to compute the magnetization $m_0$ from
the cavity Hamiltonian (\ref{eq:H^cav}).  In this ' cavity-mean-field
approximation ', one approximates the cavity Hamiltonian acting on spin $0$ by \begin{equation}
H_{0}^{cav-MF}=-\xi_{0}\sigma_{0}^{z}-\sigma_{0}^{x}\sum_{i=1}^{K}J_{0i}\langle\sigma_{i}^{x}\rangle\label{eq:H^cav-MF}\end{equation}
 This implies that $B_{0}=\sum_{i=1}^{K}J_{0i}\langle\sigma_{i}^{x}\rangle$,
giving the explicit recursion equation relating the $B$ fields: \begin{equation}
B_{0}=\sum_{i=1}^{K}J_{0i}\frac{B_{i}}{\sqrt{B_{i}^{2}+\xi_{i}^{2}}}\tanh\beta\sqrt{B_{i}^{2}+\xi_{i}^{2}}\ .\label{eq:mapping_Kfinite}\end{equation}

This recursion induces a self-consistent equation for the distribution
$P(B)$ which can now be written explicitly: 
\begin{equation}
P(B)=\int \prod_{i=1}^K\left[dB_{i}d\xi_idJ_{0i}P(B_{i})\rho(J_{0i})\pi(\xi_i)\right]
\delta\left(B-\sum_{i=1}^{K}\frac{J_{0i}B_{i}}{\sqrt{B_{i}^{2}+\xi_{i}^{2}}}\tanh\beta\sqrt{B_{i}^{2}+\xi_{i}^{2}}\right)\
.\label{eq:PdeB_Kfinite}
\end{equation}
  
\section{A simple test of the projected cavity recursion: the pure
  ferromagnet}
\label{se:ferro}
It is useful to test the validity of the two levels of
approximation described above (the projected cavity mapping and the
cavity-mean-field approximation) versus precise results using the full numerical
sampling of Suzuki-Trotter trajectories. This can be done  in the case
of a pure ferromagnet in a uniform transverse
field, defined by $\rho(J)=\delta(J-1)$ and
$\pi(\xi)=\delta(\xi-h)$. We thus study the problem defined by  the Hamiltonian \begin{equation}
H_{PF}=-h\sum_{i}\sigma_{i}^{z}- \sum_{(ij)}\sigma_{i}^{x}\sigma_{j}^{x}\ \label{H_C}
\end{equation}
on a random regular graph where each site has $z=K+1$ neighbours. The
exact cavity study in terms of continuous-time spin trajectories has
been done in the case $K=2$ by~[\cite{Krzakala}], we shall use
their results as a benchmark in order  to test the accuracy of the three levels of approximations:

{\textbf {1: Naive mean field}}: The Eq.(\ref{eq:BCS_equation}) for the
spontaneous field $B$ in the $x$ direction becomes:
\begin{equation}
B=(K+1) \frac{B}{\sqrt{h^{2}+B^{2}}}\tanh(\beta\sqrt{h^{2}+B^{2}})\
.
\label{eq:BCS_equation_ferro}
\end{equation}

{\textbf {2: Cavity-mean-field}}: The recursion relation
(\ref{eq:mapping_Kfinite}) now becomes a self-consistent equation for the
spontaneous field $B$ in the $x$ direction (which is site-independent):
\begin{equation}
B=K \frac{B}{\sqrt{h^{2}+B^{2}}}\tanh(\beta\sqrt{h^{2}+B^{2}})
,
\label{eq:mapping_Kfinite_ferro}\end{equation}
it differs from the naive mean field only by the replacement $K+1\to
K$.

{\textbf {3: Projected cavity mapping}}:
The system of spin $0$ and its $K$ neighbors is  described by
the local Hamiltonian
\begin{equation}
H_{0}=-h\sigma_{0}^{z}-\sum_{i=1}^{K}\left(h\sigma_{i}^{z}+B\sigma_{i}^{x}+\sigma_{0}^{x}\sigma_{i}^{x}\right)\label{eq:H^cav_ferro}
\end{equation}
The self-consistent equation for  the
spontaneous field $B$ in the $x$ direction is given by
\begin{equation} 
m_0=\frac{Tr(\sigma_{0}^{x} e^{-\beta H_0})}{Tr
e^{-\beta H_0}}=\frac{B}{\sqrt{h^2+B^{2}}}\tanh\beta\sqrt{h^{2}+B^{2}} \ .
\label{scFerro}
\end{equation}
These traces involve matrices of size $2^{K+1}$, they are easily computed numerically when $K$ is not too large.

\begin{figure}
 \centering
 \includegraphics[scale=0.3,angle=-90]{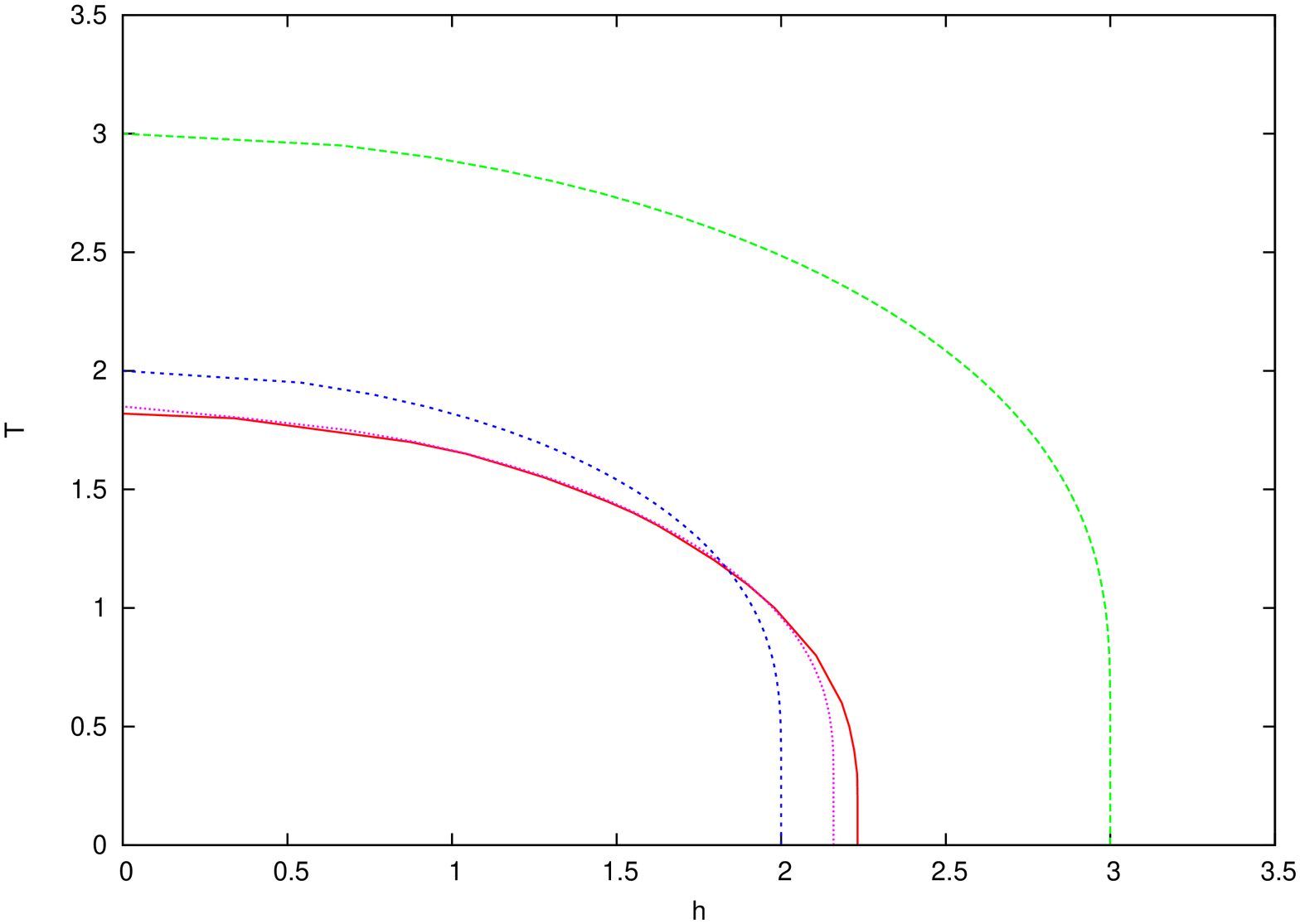}

\includegraphics[scale=0.65,angle=0]{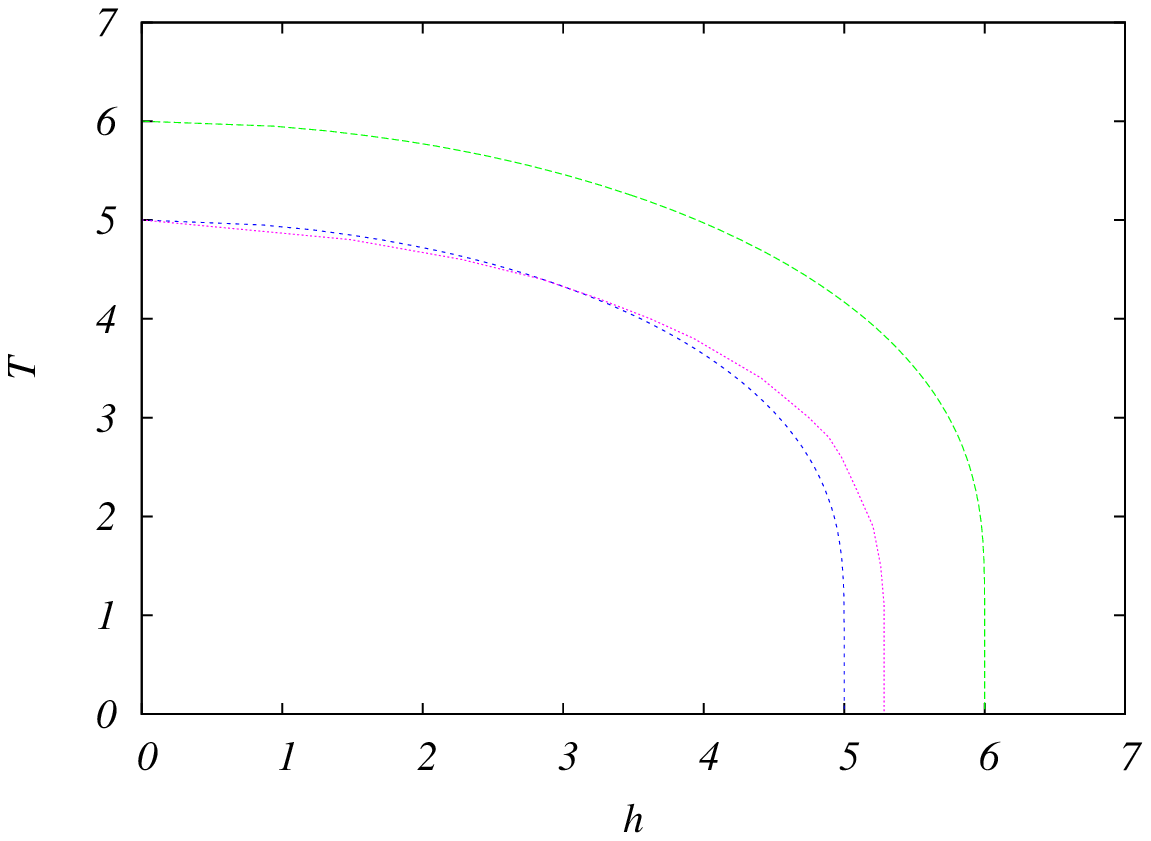}
  \caption{\textbf{Top:} The phase diagram of the ferromagnet in a transverse field,
    on a Bethe lattice with $K+1=3$ neighbours per spin. The plot
    shows the critical temperature versus transverse field, separating
    the low-$T$ ferromagnetic region from the high-$T$ paramagnetic
    one.
The full red curve
  is the exact result obtained by continuous time spin trajectory
  population dynamics~\cite{Krzakala}. The green long-dashed curve is
  the naive mean field result. The blue dashed line is the result from
  the cavity-mean-field approximation. The dotted purple curve is the  result
  obtained from the projected cavity mapping.
\textbf{Bottom:} Same plots with $K+1=6$ (without the exact result), showing that 
 the approximation going from the projected cavity mapping to the
 cavity-mean-field improves at larger $K$, as expected, and becomes
 rather accurate already for this moderate value of $K$. }
\label{fig:compare}
\end{figure}

The phase diagram obtained with these three approximations is
plotted in Fig.\ref{fig:compare}, and compared to the exact one of~[\cite{Krzakala}].
The projected cavity mapping gives a rather precise approximation of
the phase diagram, much better than the one obtained by naive mean
field or by the well known static approximation (see~[\cite{Krzakala}]).
A slightly better result can be obtained by using the parameterized form (\ref{psi_proj}) variationally inside the Bethe free energy~\cite{Zamponi}.

One may also use the cavity method as a mean field approximation to
study the finite dimensional problem. It turns out that it is able to
locate the transition point rather accurately. For a three dimensional
case, the simulations through a cluster Monte-Carlo method of the
Suzuki-Trotter formulation~\cite{Blote}  gives a  zero-temperature transition
occuring at $h_c=5.16$. The projected cavity mapping (with $K=5$) gives
$h_c=5.28$, and the cavity-mean-field gives $h_c=5.00$. In two
dimensions, the numerical result of~[\cite{RiegerKawashima,Blote}] is $h_c=3.04$, the projected cavity mapping (with $K=3$) gives
$h_c=3.22$, and the cavity-mean-field gives $h_c=3.00$.

\section{Phase diagram of the random transverse-field ferromagnet}
\label{RTFF}
Let us apply the cavity-mean-field approximation to the random
transverse-field ferromagnet. The whole problem reduces to finding the
distribution of fields $P(B)$ which solves the self-consistent
equation~(\ref{eq:PdeB_Kfinite}). It turns out that the solution of this equation is
more subtle than  the one of seemingly similar equations obtained in the classical
case.

In order to understand the properties of this solution, it is useful to
go back to the cavity mapping (\ref{eq:mapping_Kfinite}). Let us  iterate
this mapping $L\gg 1$ times on the Bethe lattice.
For $L$ finite and $N\rightarrow\infty$ the corresponding graph
is just a rooted tree with branching factor $K$ at each node and
depth $L$. The field $B_{0}$ at the
root is a function of the $K^{L}$ fields on the boundary. In order
to see whether the system develops a spontaneous ferromagnetic order, we study the value of
$B_{0}$ in linear response to infinitesimal fields $B_{i}=B\ll1$
on the boundary spins. This is given by 
\begin{equation}
B_{0}/B=\Xi\equiv\sum_{P}\prod_{k\in
  P}\left[\frac{J_k\tanh(\beta\xi_{k})}{\xi_{k}}\right]\
.\label{eq:Xi_def}
\end{equation}
 where the sum is over all paths going from the root to the boundary,
and the product $\prod_{k\in P}$ is over all edges along the path
$P$. 

\subsection{Phase transition and directed polymers}

The response $\Xi$ is nothing but the partition function for
a directed polymer in a random medium (DPRM) on a tree, where there exists on each edge $k$
of
the tree a random energy  $E_k$ given by
$e^{-E_{k}}=J_k(\tanh(\beta\xi_{k})/\xi_{k})$ and the temperature of
the polymer
has been set equal to one. The random energies are independent
identically distributed random
variables. We need to compute the
large $L$ behavior of $\Xi$. As $\Xi$ is a partition function, this
is naturally characterized
by the free energy
\begin{equation}
\Phi=\lim_{L\to\infty}(1/L)\ln\Xi\ . 
\end{equation}
If $\Phi<0$ the effect of a small boundary field decays with distance:
the spin system is paramagnetic. If $\Phi>0$ it is ferromagnetic.

The general method for computing $\Phi$ in DPRM on trees
has been developed by Derrida and Spohn~\cite{DerrSpohn}.
Their solution can be expressed in terms of the convex function
 \begin{equation}
f(x)=\frac{1}{x}\ln\left[K\int dJ \rho(J) \; \int d\xi \pi(\xi)
  \;\left(\frac{J \tanh(\beta\xi)}{\xi}\right)^{x}\right]\
.\label{eq:f(x)gen}
\end{equation}
 Let us denote by $x=m$ the value of $x\in[0,1]$ where this function
 reaches its
minimum.The free energy of the DP is then
$\Phi=f(m) $. The phase transition line in the $(h,T)$ plane separating
the ferromagnetic from the paramagnetic phase is thus obtained by
solving the equation $f(m)=0$.

\begin{figure}
      \begin{overpic}[scale=1.9,unit=1mm]%
            {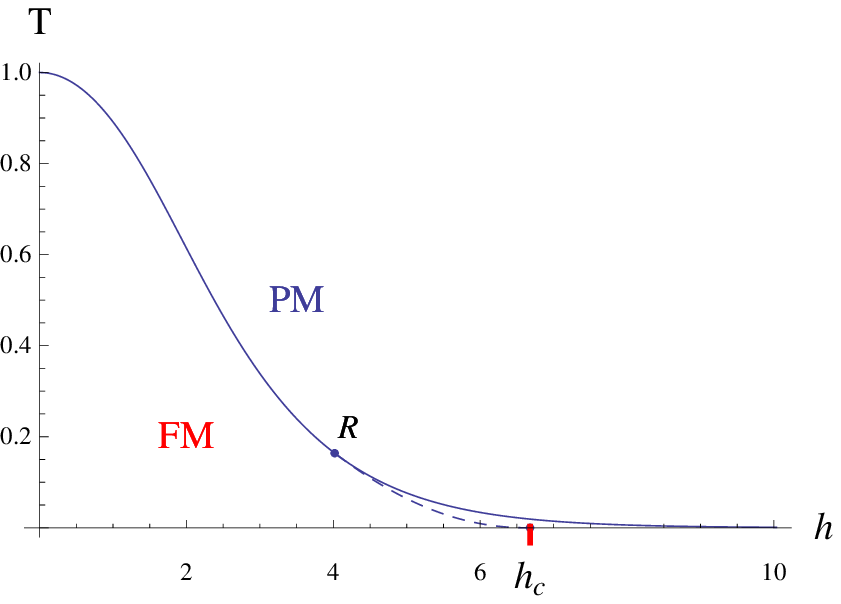}
        \put(65,49){\includegraphics[scale=1.07]%
            {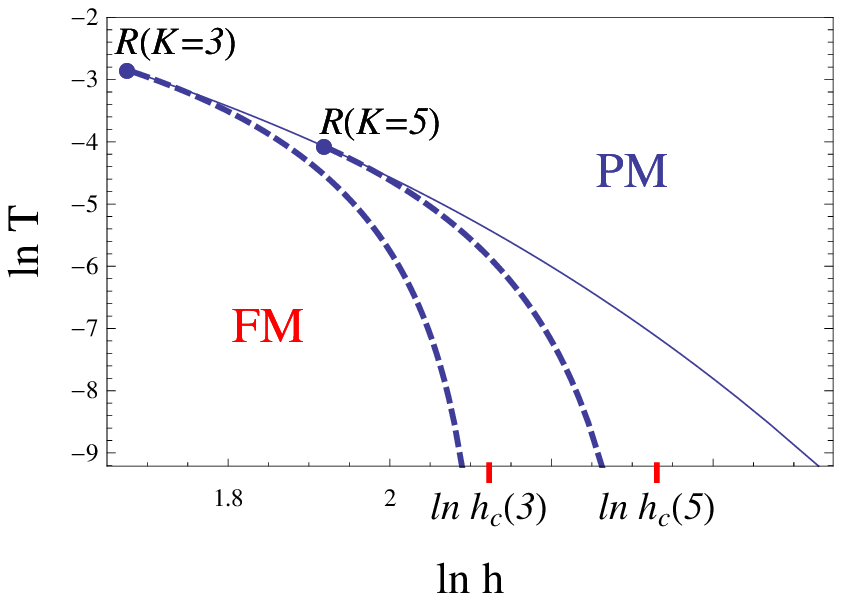}}
      \end{overpic}
      \caption{Phase diagram of the quantum ferromagnet in a random
        transverse field, as found with the RS cavity method,  in the
        plane of the disorder strength $h$ and the temperature $T$.
      The RS transition line (solid) is determined by the condition on the convex function: $f(x=1,T,h)=0$, 
      and separates
      the ferromagnetic (FM) state ($f(x=1,T,h)<0$) from the paramagnetic (PM) state  ($f(x=1,T,h)>0$).
      The point $R$ on the RS line corresponds to the minimum of the convex function: $f'_x(x=1,T,h)=0$. 
      It is the beginning of the RSB transition line, determined by the minimum of the convex function 
      in the RSB ($x^*<1$) phase:
     $f(x^*,T,h)=0$ and $f'_x(x^*,T,h)=0$. The inset shows details of
     the RSB lines in the cases $K=3$ and $K=5$. The critical values of disorder $h_c(K)$, at which the 
      zero-temperature quantum phase transition to the paramagnetic phase happens,
     for the two connectivity numbers $K$ are:
  $h_c(K=3)=8.36$ and $h_c(K=5)=10.28$. }
\label{fig:PhaseDiagram}
      \end{figure}

\begin{figure}
      \includegraphics[scale=1.07]%
            {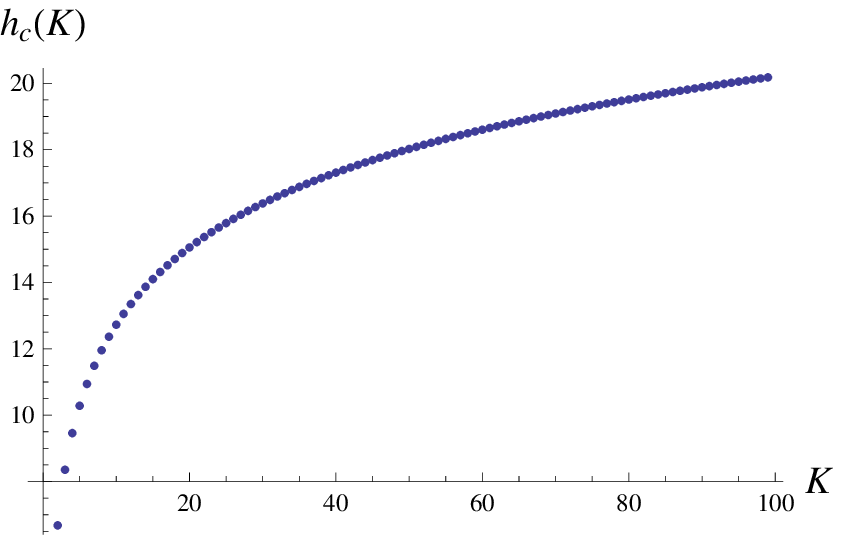}
      \caption{The critical value of disorder $h_c(K)$,
      as a function of the connectivity number $K$. 
    }
\label{fig:hcK}
      \end{figure}

In the case of the random transverse-field ferromagnet one obtains
\begin{equation}
f(x)=\frac{1}{x}\ln\left[\frac{K^{1-x}}{1+x}\right] +\log\frac{2}{T}+
\frac{1}{x}\log\left[\frac{T}{h}\int_0^{h/T} du \left(\frac{ \tanh(u)}{u}\right)^{x}\right]\
,\label{eq:f(x)gen2}
\end{equation}
from which one deduces the following phase diagram (see Fig.\ref{fig:PhaseDiagram}).:

\begin{itemize}
\item
When the field $h$ is smaller than a value $h_R$ the phase
transition line is given by $f(1)=0$. This gives back the naive mean
field result of (\ref{Tc_MF}): $h=\int_0^{h/T_c} du \tanh(u)/u$.
The value $h_{R}$, and the corresponding value $T_R=T_c(h_R)$, are obtained by
solving the pair of equations $f(1)=0,\ f'(1)=0$.
\item
When $h>h_R$, the phase transition is given by solving the equations
$f(m)=0,\ f'(m)=0$. In this regime the solution is at $m<1$. This
low temperature, large disorder part of the phase transition line
departs from the naive mean field prediction. In particular, at
variance with the naive mean field result, one finds
that there exists a zero temperature quantum phase transition at a
critical disorder $h_c$. This is obtained using the zero-temperature
limit of $f$:
\begin{equation}
f_0(x)=\frac{1-x}{x}\ln K -
\frac{1}{x}\log(1-x^2)-\log\left(\frac{h}{2}\right)
,\label{eq:f0}
\end{equation}

and solving the two equations $f_0(m)=f_0'(m)=0$.
Fig.\ref{fig:hcK} shows $h_c$ as function of $K$. In the large $K$ limit
$h_c$ diverges, giving back the naive mean field
result. However this divergence is very slow: 
one finds that $m\sim 1-1/\log K$ and $h_c\sim e \log K$.

\end{itemize}

The  phase transition line has two parts: a small-disorder
part on the left of the point $R$ with coordinates ($h_R,T_R)$, and a
large-disorder part on the right of this point. The properties of these two regimes of
the transition are very different. In order to understand this
difference, it is useful to give a closer look at the DPRM problem
using the replica method, as was first done in~[\cite{CookDerrida90}].
The DP partition function $\Xi$, defined in (\ref{eq:Xi_def}), depends
on the random quenched variables $J_n,\xi_{n}$. To compute the value
of $\log\Xi$ for a \emph{typical} sample one needs to do 
a quenched average of $\log\Xi$ over these random
variables, denoted by $\overline{\log\Xi}$. 
As a first step one might try to approximate it by the annealed
average $\log\overline\Xi$. This gives precisely the naive-mean field result.
To go beyond this `annealed approximation' one can introduce replicas, using
$
\overline{\ln\Xi}  =  \lim_{n\rightarrow0}(\overline{\Xi^{n}}-1)$.
 The average of $\Xi^{n}$ is obtained by a sum over $n$ paths, \begin{eqnarray}
\overline{\Xi}^{n}=\sum_{P_{1},\dots,P_{n}}\overline{\prod_{a=1}^n\left[\prod_{k\in P_a}\left(J_k\frac{\tanh(\beta\xi_{k})}{\xi_{k}}\right)\right]}\ .\label{xinav}\end{eqnarray}
 
One can then proceed by testing various hypotheses concerning the
structure of the paths which dominate the replicated partition
function~(\ref{xinav}):
\begin{itemize}
\item In the so-called replica symmetric (RS) hypothesis one assumes that the leading contribution to (\ref{xinav})
comes from non-overlapping paths. This gives
\begin{equation}
\overline{\Xi^{n}}=\exp(Ln f(1))=\left(\overline{\Xi}\right)^{n}\ .\end{equation}
The RS hypothesis is equivalent to the annealed approximation in this
case. 
\item
The so-called one step replica-symmetry-breaking  (1RSB) hypothesis assumes that the leading contribution to (\ref{xinav})
comes from patterns of $n$ paths which consist of $n/m$ groups of
$m$ identical paths, where the various groups go through distinct
edges. This gives: \begin{equation}
\overline{\Xi^{n}}=\exp(Ln f(m))
\end{equation}
 where $f(x)$ is the function introduced in (\ref{eq:f(x)gen2}). In
the replica limit $n\rightarrow0$, the parameter $m$ should belong
to the interval $[0,1]$. It turns out that one should minimize the function $f(m)$ over $m\in[0,1]$
(the fact that one should minimize $f$, and not maximize it as one
would have thought naively, is a
well-known aspect of the replica method~\cite{MPV}).
The low-temperature large-disorder part of the phase transition line
corresponds
to this 1RSB solution.
\end{itemize}

\subsection{Replica symmetry broken phase of the polymer and Griffiths
  phase of the ferromagnet}
This replica solution of our DPRM, with a RS phase at
small disorder and a 1RSB phase at large disorder, fully agrees with
the original travelling wave solution of~[\cite{DerrSpohn}].
So the point R on the paramagnetic-ferromagnetic transition line of the quantum random
transverse-field ferromagnet corresponds to a 1RSB transition in the
auxiliary DPRM problem. This is a glass transition of the polymer, 
similar to the one found in the random energy
model~\cite{DerridaREM,GrossMezard}, where the measure on the paths
condenses on a finite number of paths. A useful order parameter to
describe this transition is
 \begin{equation}
Y=\frac{1}{\Xi^2}\sum_{P}\prod_{k\in
  P}\left[\frac{J_k\tanh(\beta\xi_{k})}{\xi_{k}}\right]^2.
\end{equation}
This is nothing but the sum $\sum_{P}w_{P}^{2}$
where $w_{P}$ is the relative weight of path $P$ in the measure
(\ref{eq:Xi_def}). Its inverse gives a measure of the effective number
of paths contributing to $\Xi$. 
In the RS phase one finds that $Y$ vanishes in a
typical sample. This can be understood from the fact that the DPRM
partition
function gets a contribution from a large number of paths (diverging
in the large $L$ limit). In fact the DPRM problem has a finite entropy
in this RS phase. On the contrary, in the 1RSB phase one finds that
$Y$ is finite and fluctuates strongly from sample to sample. For
instance, its first moments are given by~[\cite{MPSTV85}]:
\begin{equation}
\overline Y=1-m\ \ \ ; \ \ \ \overline {Y^2}=
\frac{1}{3}(1-m)+\frac{2}{3}(1-m)^2,
\end{equation}
and the entropy density of the polymer vanishes in this whole 1RSB
phase.

The DPRM problem is an auxiliary construction used to study
the phase transition in the original problem of the random transverse field ferromagnet. So
the natural question is: what are the consequences of the 1RSB
transition in the DPRM concerning the original quantum spin system?
Let us go back to (\ref{eq:Xi_def}), which connects the two
problems. We consider a large tree of length $L\gg 1$. The condensation
of the paths in the DPRM means that, if a small field is applied to
the boundary,
the value of the effective field
$B_{0}$ on the root is dominated by a finite number of paths to the
boundary.
So the root spin feels the effect of only a finite number
of the boundary spins. The physics is dominated by rare events: rare
boundary spins have a dominant influence of the central spin, and
it is building up on these rare events that the system can or cannot
develop a long range ferromagnetic order. This is very similar to the 
Griffiths phase, found originally in the one-dimensional case~\cite{Fisher1992}, and argued to be present at least in
low-dimensional systems. It is interesting to see that we find here a similar
rare-event dominated phase, using the mean-field type Bethe
approximation
which is valid in the opposite limit of large connectivity, $K\gg
1$. The next subsections will give a more detailed discussion of the
connections between the results obtained with the cavity method and
those known, or conjectured, for finite-dimensional problems. Let us
first summarize the main observables that are accessible using the
cavity method.

One can study the practical consequences effect of the rare events
on both sides of the phase transition, in the neighborhood of the
phase transition line, on the right of point R. The study of~[\cite{FeiIofMez2010}]  has discussed these effects in the context of the
superconductor-insulator transition. It is easy to translate them to
the present situation of the ferromagnetic-paramagnetic transition.
\begin{itemize}
\item
In the ferromagnetic phase the order parameter (here we use
as order parameter the typical local magnetic field  $B_{typ}$ in the z
direction, obtained as $B_{typ}=\exp(\overline {\log Z})$) has an anomalous scaling and
distribution. Imagine one enters the ferromagnetic phase at a given
temperature $T$ by varying the disorder $h$. 
At large enough temperature the scaling is $B_{typ}\sim\sqrt{h_c-h}$, as one expects in
a mean field system. Then lowering the temperature there is a region
above $T_R$ where the scaling is  $B_{typ}\sim(h_c-h)^{a(T)}$ . The
exponent $a(T)$ diverges when $T\to T_R$. In the 1RSB regime, i.e. for
$T<T_R$, the typical field has an essential singularity in
$e^{C/(h_c-h)}$. In this regime the distribution of the local order
parameter has a power law tail $B_{typ}^m/B^{1+m}$. As $m<1$, the average
order parameter is divergent.
\item 
In the paramagnetic phase the local susceptibility fluctuates
strongly from site to site and nonlinear effects are crucial.
Consider the system in a very small uniform magnetic field $B_{ext}$ and look at
the local order parameters $B_i$ induced by this external field. The
typical value of $B_i$ scales linearly with $B_{ext}$, but the moments $\overline{B^{x}}$
with $x>m$, and in particular the mean $\overline{B}$, are divergent
at the level of the linear response to $h$: they behave non-linearly,
as $\overline{B^{x}}\sim C B_{ext}^{m}$. The distribution of the local
susceptibilities $\chi_i=B_i/B_{ext}$ has a power law tail
$P(\chi)\sim C/\chi^{1+m}$.
\end{itemize}

\subsection{The one-dimensional case}
The one-dimensional RTFF is the best understood case. The strong
disorder decimation procedure of Ma-Dasgupta-Hu, developed by Fisher~\cite{Fisher1992}, becomes exact at the transition point because of the
existence of an infinite disorder fixed point. It  can also  be checked versus
alternative methods like the mapping of the problem to free fermions~\cite{RiegerYoung,IgloiRieger98}, and the exact solution of McCoy and Wu~\cite{MacCoyWu}. A large corpus of results has been obtained by
these methods, concerning the scaling behaviour and the Griffiths
phase. The strong disorder decimation has turned out to be useful in a
broad range of problems~\cite{Monthus}.

Let us mention only some aspects of the physics of the one-dimensional case
for which one can draw a
comparison between the exact results and those of the cavity method.
The exact analysis of the problem has shown that the zero-temperature critical point is
located at the width of the random transverse field $h=h_c$ such that $\overline {\log
\xi}=\overline{\log J}$, or explicitely: $\int d\xi \pi(\xi) \log
\xi=\int dJ \rho(J)\log J$. In our case this gives $h_c=2$.
At the
critical point the system  displays activated dynamical scaling: the scaling
exponent $z_c$, relating the characteristic time scale $\tau$ to the
characteristic length scale $\xi$ through $\tau=C \xi^{z_c}$, is formally
infinite. Instead the scaling is $\log \tau= C \xi^\psi$. 
The Griffiths phase can be studied close to the
critical point. Defining $\delta=(\overline {\log h}-\overline {\log
J})/(var[\log h]+var[\log J])$, where the overline denotes the average
and $var$ denotes the variance, this exponent has been shown to 
behave as $z(\delta)\sim 2/\delta +C$, where $C$ is a non-universal
constant. This exponent $z(\delta)$ also appears in the singular
response of the system to an external magnetic field $H$: in the
paramagnetic phase, the linear
susceptibility diverges, and the magnetization behaves as $M\sim
|H|^{1/z(\delta)}$.

Let us now present the results of the cavity method described in the previous
subsection. The one-dimensional case is obtained by taking
$K=1$. Actually it is useful to keep $K=1+\epsilon$ at intermediate
steps of the computation, and take the $\epsilon\to 0$ limit in the
end. At zero temperature, the minimum of the function $f(x)$ is
obtained at $x=m=\sqrt{\epsilon}$, and the critical value of $h$,
obtained by solving $f(m)=0$, is $h_c=2(1+2\sqrt{\epsilon})$. When
$\epsilon\to 0$, $h_c$ goes to $2$ which is the exact result, and $m\to
0$. The exponent $m$ is the one that controls the decay of the local
field distribution (which goes as $B^{-(1+m)}$), and the distribution
of local susceptibilities. From the analysis of~[\cite{FeiIofMez2010}],
this exponent $m$ controls the response to a uniform external field,
$M\sim | H| ^m$. So one should identify $m=1/z(\delta)$. At the
critical point we have found $m\to 0$. This agrees with $z_c=\infty$
and the activated scaling behaviour. Let us now study the behaviour
close to the critical point, at $h=h_c(1+y)$. At $K=1$ (or
$\epsilon=0$), one obtains $m=y+ O(y^2)$. So the prediction of the
cavity method is $z=1/y$. It turns out that, when expressed in terms
of the distance to the critical point $\delta$, this gives the exact result $z=1/(2\delta)$.

The quantum cavity method is exact in one dimension, if one uses the full
mapping of spin trajectories given in (\ref{st_update}). It must thus
reproduce the exact results. What we have seen is that even the
simpler cavity-mean-field approximation gives a fair description of
the problem, including the exact location of the critical point, the
activated dynamical scaling, and the correct value of $z$ close to the
critical point. Actually it gives the exact correlation even for a
finite length problem, as we now show. 

Consider a chain described by the Hamiltonian
\begin{equation}
H_{1d}=-\sum_{n=1}^L\xi_{n}\sigma_{n}^{z}-\sum_{n=1}^{L-1}J_{n}\sigma_{n}^{x}\sigma_{n+1}^{x}\ .\label{H_1d}
\end{equation}
In order to see the onset of long-range order, one can fix that the
end-spin $\sigma_L$ is in the eigenstate of $\sigma_L^x$ with
eigenvalue $+1$, fix $\xi_L=0$, and compute the so-called 'surface' magnetization at
the other end-point in the ground state $m_1=\langle 0 \vert
\sigma_1^x\vert0\rangle$. By using the Jordan-Wigner transformation of
this problem to free fermions, Igl\'oi and Rieger have shown
that~\cite{IgloiRieger98}:
\begin{equation}
m_1=\left[1+\sum_{n=1}^{L-1}\left(\prod_{r=1}^n\frac{\xi_r}{J_r}\right)^2\right]^{-1/2}
\label{m1exact}
\end{equation}
This formula, which is exact for a finite system, can then be used to
study the critical behavior of the problem. For instance,~[\cite{IgloiRieger98}] shows how to deduce from this formula several
interesting physical properties:   the typical and the average surface
magnetization differ,  the average critical magnetization scales like
$L^{-1/2}$, the exponent $\beta_s$ giving the surface magnetization
close to the critical point is equal to $\beta_s=1$, the correlation length exponent exponent $\nu$ is equal to
$\nu=2$. 

Let us now use the cavity-mean-field method on this one-dimensional
problem. At zero temperature, formula (\ref{eq:mapping_Kfinite})  gives the recursion
\begin{equation}
B_{n-1}=J_{n-1}\frac{B_n}{\sqrt{B_n^2+\xi_n^2}},
\label{mapd1}
\end{equation}
which should be initialized with $B_L=\infty,\ \xi_L=0$. It is easy to
see that, iterating the mapping~(\ref{mapd1}), and computing
$m_1=B_1/\sqrt{B_1^2+\xi_1^2}$, one obtains the exact formula
(\ref{m1exact}). Therefore the cavity method, even using the simplified,
mean-field mapping, gives the exact value of the surface
magnetization, also for a finite system. Its predictions thus agree
with the known results, including the values of the exponents
$\beta_s$ and $\nu$.

\subsection{Dimensions two and above}

There is no exact solution of the  two-dimensional RTFF problem. 
This problem has been studied mostly using the strong-disorder
decimation procedure~\cite{Motrunich,KovacsIgloi_prb2010,KovacsIgloi10}, and also using cluster Monte-Carlo simulations
of the 2+1 dimensional problem obtained in the Suzuki-Trotter
representation\cite{Pich,RiegerKawashima}. These works have mostly used a different
normalization, in which $\rho(J)$ is the uniform distribution on
$J\in [0,1]$, and $\pi(\xi)$ is uniform on $[0,h]$. In this subsection
we shall thus adopt this other normalization. The results
obtained both by decimation and by Monte-Carlo indicate that the
situation is similar to the one-dimensional case. In particular, the
exponent $z_c$ is found to diverge, and $1/z(\delta)$ goes to zero as
$\delta\to 0$. In order to obtain an analytic approximation to this
two-dimensional problem, one can repeat the cavity study on the Bethe lattice
with $K=3$. 

Within the cavity-mean-field approximation, the function $f(x)$ is found to be
\begin{equation}
f(x)= \frac{1}{x} \log 3-\frac{1}{x} \log(1-x^2)-\log h.
\end{equation}
It is minimum at $x=m=.679$, and the critical value of the field,
obtained from $f(m)=0$, is
found to be $h_c=12.53$. 
This result disagrees with the strong-disorder decimation, and with
the Monte-Carlo, in two aspects: i) the value of $h_c$ (with the Monte-Carlo it
is found~\cite{Pich,RiegerKawashima} around $h_c\sim 4.2$, with the
decimation it is found~\cite{KovacsIgloi_prb2010,KovacsIgloi10,saleur}
around $h_c\sim 5.3$), ii) most importantly, the fact that the value of
$m$ found in the cavity method does not
go to zero at the transition.

\begin{figure}
      \includegraphics[angle=270,scale=0.5]%
            {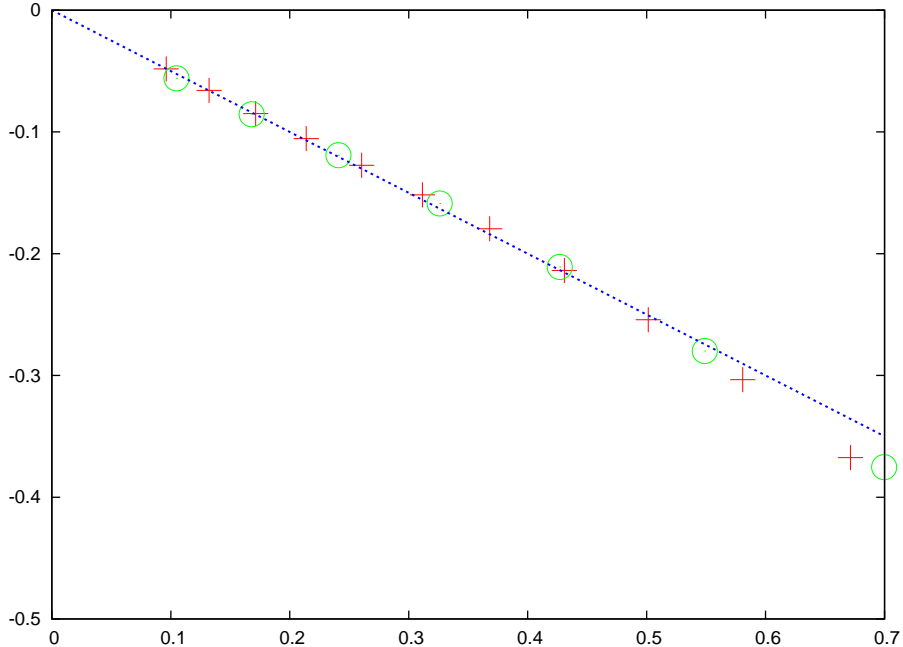}
      \caption{Phase transition in the RTFF model with $K=3$, a
        uniform distribution of couplings $J$ on $[0,1]$, and a
        uniform  distribution of fields $\xi$ on $[0,h]$.  The typical value of the local field $B$, measured as
$\exp(-\overline{\ln B})$, is plotted versus $(h_c/h)^m -1$. The
crosses are obtained with the  cavity-mean-field approximation and the
choice $h_c=12.53$ obtained from the analytic study of the DPRM
problem, which predicts a linear dependence $ \exp(-\overline{\ln B})=C[(h_c/h)^m -1]$. The circles are obtained  with the projected cavity mapping,
and a fitted value $h_c=7.5$. In both cases the exponent $m$ is the
one found from the DPRM analysis, $m=0.679$. The straight line is a
guide to the eye.
    }
\label{fig:K3d2}
      \end{figure}

One may wonder to what extent these results depend on the
approximation. If instead of doing the cavity-mean-field approximation
one uses the more accurate projected cavity mapping, the value of
$h_c$ is changed to $h_c\sim 7.5$, but the value of $m$ is
unchanged. This is seen from Fig.\ref{fig:K3d2}, which shows 
the inverse of the typical value of the local field $B$, measured as
$\exp(-\overline{\ln B})$, plotted versus $(h_c/h)^m -1$. The
prediction of~[\cite{FeiIofMez2010}] is that the inverse of the typical
value of $B$ should go to zero linearly in $(h_c/h)^m -1$. Both data
sets, obtained from the cavity-mean-field approximation and from the
projected cavity mapping, find this behaviour, but the values of $h_c$
in both data sets are distinct (respectively $h_c=12.53$, as predicted
from our analytic study with the DPRM, and $h_c=7.5$, found from the fit).

The conclusion is that, while the non-universal critical value $h_c$
depends quite a lot of the approximation used in the cavity method,
the value of $m$ does not, and the statement that $m \neq 0$ at
the transition seems to be a solid result for the problem on the Bethe
lattice with any connectivity $K+1\geq 3$. Notice that this result still
implies that there is a strong hierarchy of local magnetic fields: the
average field diverges, and so does the average susceptibility. But it
is not an infinitely strong hierarchy as happens when $m=0$.  It
should be noticed that our best cavity analysis, using projected
cavity mapping, still involves some
approximation, and it would be interesting to carry out the full
Suzuki-Trotter trajectory mapping in order to get the exact solution
for the Bethe lattice. At present, our approximate treatment gives $m>0$ at the transition for the Bethe
lattice problem  a result which disagrees with a recent study
from strong disorder decimation~\cite{KovacsIgloi10}. It should be
noticed that our cavity approximations shoud become exact in the large
$K$ limit of the Bethe lattice. Assuimng that the result
$z_c\neq\infty$ is correct for the Bethe lattice, this raises the
question of the existence and value of a critical dimension above
which the finite dimensional problems have $z_c\neq\infty$.

It would thus be interesting to revisit 
the numerical results in $d=2$ and $d>2$, and on the Bethe lattice, giving $m=0$ at the
transition. Concerning the Monte-Carlo approach, let us point out in
particular that the RSB effects happen at very low temperature, but
they completely change the physics of the problem (for the problem
with $K=3$ and uniform couplings on $[0,1]$, we have found with the
cavity-mean-field approximation that the
RSB transition point on the ferromagnetic-paramagnetic transition line
occurs at $T_R\sim 0.087$). So the numerical results
obtained by extrapolation from a not-very-low temperature to $T=0$
might be questionable.

\section{Summary}
\label{se:conc}
We have proposed here a well defined mean-field scheme to study
quantum ferromagnets in random transverse fields. This scheme consists
in studying the problem on a random regular graph with fixed
connectivity $K+1$. The problem at finite $z$ is qualitatively distinct
from the infinite $z$ case, described by the naive mean
field theory. It has a zero temperature quantum phase transition at a
finite critical value of the width of the distribution of random
transverse fields.

The phase diagram has been studied by the RS quantum cavity method. A
full study would involve a complicated population dynamics in terms of
imaginary time spin trajectories, which would be extremely costly in
terms of computing efforts. Instead we have resorted to a simple
approximation, the cavity mean-field approximation, which
parameterizes the imaginary time spin trajectories by a single number,
a local cavity longitudinal field. Within this approximation, the RS
cavity method becomes a simple recursion relation on these local
fields.
While this recursion looks structurally similar to the one found in
classical spin systems, its physical content is very different. When
one iterates $L$ times the linearized
recursion describing the phase transition, one obtains the partition
function
for a classical directed polymer in a random medium on a tree. This 
problem has a glass transition. The small disorder phase of the
polymer describes the high temperature  part of the
ferromagnetic to paramagnetic phase transition, which is identical to
what one gets in the naive mean field approach. The  large disorder phase of the
polymer describes the lower temperature  part of the
ferromagnetic to paramagnetic phase transition, which has a completely
different behavior. This glass phase of the polymer, which displays
replica symmetry breaking, actually describes the Griffiths
phase of the quantum spin system in the neighborhood of its quantum
phase transition. The results obtained with the cavity analysis fully
agree with the known behaviour of the system in one dimension. On
Bethe lattices with connectivity $K+1\geq 3$, we find a Griffiths phase,
as in one dimension, but the critical point is not an infinite disorder
fixed point: the exponent $m$ which governs the tail of the local
field distribution does not vanish at the transition, meaning that the
critical exponent $z$ does not diverge. it would be interesting to
know if there is a critical dimension of finite dimensional problems
beyond which such a behaviour appears.

One should be aware of the fact that we have been using here a RS
cavity method for the quantum spin system. Nevertheless, even within
this RS method, the cavity equations map onto a classical problem
which exhibits RSB. It would be very interesting, in particular when
applying this formalism to spin glasses, to study the effect of RSB
within the quantum cavity method itself.

\section*{Acknowledgments}
 The results in Sect.\ref{RTFF} are straightforward consequences of
 the works~[\cite{IoffeMezard2010,FeiIofMez2010}] in collaboration  with L. Ioffe and
 M. Feigelman. MM wants to thank them, and also G. Biroli, D. Huse,
 G. Semerjian, M. Tarzia and F. Zamponi for useful exchanges. 
The $K=2$ graph in Fig.~\ref{fig:compare} was originally computed by
F. Zamponi, who has kindly given us the data of the exact result
obtained by the continuous time spin trajectory
  population dynamics. The grant 
Triangle de la Physique 2007-36 has supported the work of OD and our
collaboration with L. Ioffe.

\end{document}